\documentclass[runningheads,a4paper]{llncs}

\usepackage[T1]{fontenc}
\usepackage{graphicx}
\usepackage{amsmath,amssymb}
\usepackage{booktabs}
\usepackage{multirow}
\usepackage{microtype}
\usepackage{url}
\usepackage[hidelinks]{hyperref}
\hypersetup{
  pdftitle={Predictive Enhancement Calibration for Latent Breast MRI Virtual Contrast Enhancement},
  pdfauthor={Qin Lei and Hao Wu}
}

\begin{document}

\title{Predictive Enhancement Calibration for Latent Breast MRI Virtual Contrast Enhancement}
\titlerunning{Predictive Enhancement Calibration for Breast MRI VCE}

\author{Qin Lei\inst{1,2} \and
Hao Wu\inst{1,2,3}\thanks{Corresponding author.}}
\authorrunning{Q. Lei and H. Wu}
\institute{Center for Big Data and Intelligent Medicine, The First Affiliated
Hospital of Chongqing Medical University, Chongqing, China\\
\email{qinlei@hospital.cqmu.edu.cn}
\and
Key Laboratory of Digital Health and Intelligent Medicine, Chongqing Municipal
Health Commission, Chongqing, China
\and
Chongqing Translational Medicine Center, Chongqing, China\\
\email{wuhao@cqmu.edu.cn}}

\maketitle

\begin{abstract}
Virtual contrast enhancement (VCE) synthesizes enhanced breast MR images
from pre-contrast acquisitions. Modern latent generators offer strong image
priors, but their bounded natural-image autoencoders conflict with the
non-canonical intensity scale of MRI. We show that the upper bound can alter
radiomic fidelity before generation, while scaling source and target
independently creates a coordinate inconsistency. We propose Predictive
Enhancement Calibration (PEC), which represents each pair in a shared,
case-adaptive coordinate during training and predicts its unavailable upper
endpoint from the pre-contrast image at inference. We integrate PEC with a
pretrained FLUX latent flow transformer via parameter-efficient reference
conditioning. Target round trips first isolate representation loss before
generation; near-matched conditional models then compare PEC with fixed-wide
and separate coordinates under comparable training budgets and backbone
settings. On the fixed internal MAMA100 development cohort, PEC improves all
eight point estimates in this source-only VCE setting, with paired evidence
strongest for MSE and LPIPS.

\noindent\textbf{Code:}
\url{https://github.com/tanlei0/pec-breast-mri-vce}

\end{abstract}


\section{Introduction}
\label{sec:introduction}

Dynamic contrast-enhanced magnetic resonance imaging (DCE-MRI) is central to
breast cancer detection, local staging, treatment planning, and response
assessment because contrast uptake reveals lesion vascularity beyond
unenhanced anatomy~\cite{Mann2019BreastMRI}. Gadolinium administration,
however, adds contraindications for some patients, injection-related risk,
cost, and workflow burden~\cite{Weinreb2021GadoliniumSafety}. Virtual contrast
enhancement (VCE) therefore seeks to synthesize a post-contrast image from an
available pre-contrast acquisition. The MAMA-SYNTH benchmark provides a focused
setting for evaluating whether such synthesis preserves image fidelity,
radiomic distributions, and lesion information~\cite{Osuala2026MAMASynth}.

The modelling paradigm for VCE is moving from deterministic regression toward
conditional generation. Pixel-wise regression is effective when source and
target are aligned, but its conditional-average solution can suppress
heterogeneous enhancement and fine texture. Diffusion and flow models instead
learn a conditional image distribution. At the same time, modern image
generators increasingly operate in compressed latent spaces: a pretrained
autoencoder supplies the image representation, transformer denoisers replace
convolutional U-Nets, and flow-matching objectives enable scalable
training~\cite{Rombach2022LatentDiffusion,Peebles2023DiT,Lipman2023FlowMatching,Esser2024RectifiedFlow}. These advances motivate adapting a pretrained latent
flow transformer to breast MRI instead of training a pixel-space generator
from scratch.

This adaptation exposes an intensity-coordinate problem. Conventional MRI has
no scanner-independent intensity unit, whereas a natural-image autoencoder
expects a bounded input. A fixed window provides one shared coordinate, but its
upper bound must trade clipping of the sparse enhancement tail against effective
use of the available code range. Per-image min--max scaling is adaptive, yet it
assigns different physical meanings to equal source and target pixel values.
An MRI-specific autoencoder can avoid the natural-image interface, but it
requires separate codec training and corresponding generator adaptation to a
new latent basis. These three interfaces therefore expose a design
trade-off: a fixed window preserves the pretrained basis but requires a global
range; a medical codec learns a new basis; and target-adaptive scaling preserves
detail but must remain available and consistent at inference. This leads to our
question: can the pretrained latent system be retained while its intensity
coordinate is shared, case-adaptive, and recoverable from the source?

\begin{figure}[t]
  \centering
  \includegraphics[width=\textwidth]{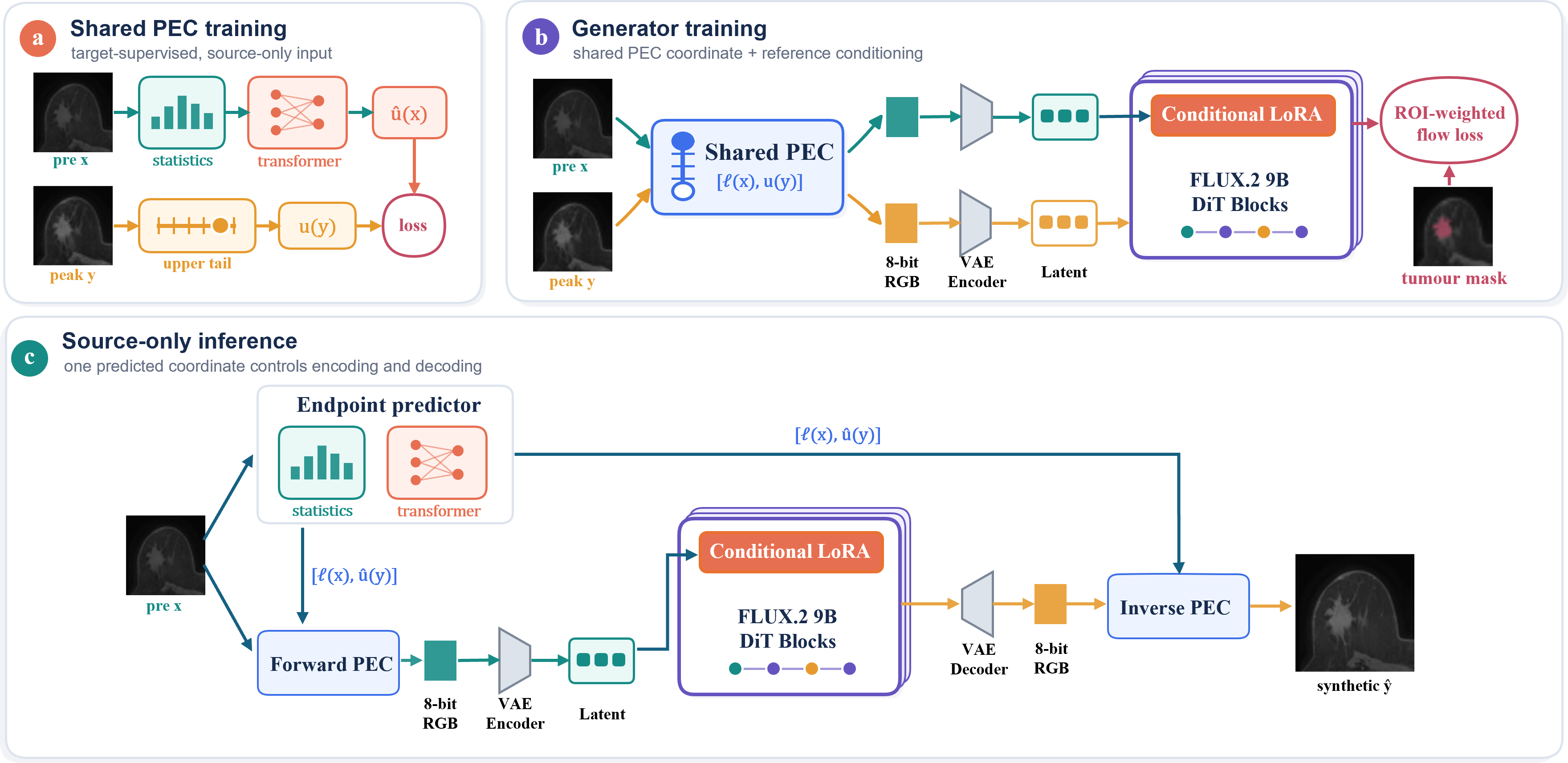}
  \caption{PEC--FLUX.2 architecture. (a) Target endpoints supervise a
  source-statistics transformer that predicts $\hat u(x)$. (b) Generator
  training maps the pre-/peak-contrast pair to the shared coordinate
  $[\ell(x),u(y)]$, encodes the resulting 8-bit RGB inputs with the frozen VAE,
  and learns conditional LoRA with an ROI-weighted flow loss. (c) Source-only
  inference reuses $[\ell(x),\hat u(x)]$ for forward PEC and output inversion;
  neither target nor tumour mask is required.}
  \label{fig:overview}
\end{figure}

We address this question with \emph{Predictive Enhancement Calibration} (PEC),
a source-predictive intensity-calibration scheme.
For each training pair, PEC derives one lower endpoint from the pre-contrast
image and one robust upper endpoint from the acquired peak post-contrast image;
both images are encoded in this shared coordinate. A compact predictor then
learns the target endpoint from source-only intensity statistics. At inference,
the predicted coordinate is used consistently for source encoding and output
decoding. We combine PEC with a pretrained FLUX.2 latent flow transformer and a
LoRA-adapted reference-token conditioning mechanism from
EasyControl~\cite{Zhang2025EasyControl}. Figure~\ref{fig:overview} summarizes the
intensity problem and the training and inference paths.

Our contributions are threefold:
\begin{itemize}
    \item We identify and measure an intensity-coordinate mismatch in latent
    breast MRI VCE, including its effect on radiomic fidelity before generation.
    \item We propose PEC, a shared source--target coordinate whose unavailable
    enhancement endpoint is predicted from the source image at inference.
    \item We integrate PEC with a large pretrained latent flow transformer using
    parameter-efficient reference conditioning and evaluate the complete system
    on the fixed internal MAMA100 development cohort.
\end{itemize}

\section{Related Work}
\label{sec:related_work}

\paragraph{Breast MRI virtual contrast enhancement.}
Early VCE systems primarily used convolutional image-to-image regression to
predict a post-contrast or subtraction image from unenhanced
MRI~\cite{Chung2023SimulatedBreastMRI}. Adversarial objectives subsequently
introduced learned realism priors, including lesion-aware and temporally
coupled synthesis~\cite{MullerFranzes2023SyntheticContrast,Osuala2025ConditionalGAN}. More recent work uses conditional diffusion to
synthesize contrast-enhanced breast MRI and latent diffusion to represent
multiple enhancement phases~\cite{Ibarra2025ConditionalDiffusion,Osuala2024ContrastKinetics}. This progression from point regression to
generative modelling addresses over-smoothing, but leaves open how quantitative
MRI intensities should enter a pretrained latent image generator.

\paragraph{Latent generative models.}
Latent diffusion reduces high-resolution generation cost by moving denoising
into an autoencoder representation~\cite{Rombach2022LatentDiffusion}. DiTs
replace the convolutional U-Net with scalable token processing, while flow
matching and rectified-flow transformers provide efficient continuous
generative trajectories~\cite{Peebles2023DiT,Lipman2023FlowMatching,Esser2024RectifiedFlow}. Domain-specific medical autoencoders offer an
alternative latent basis~\cite{Varma2025MedVAE}. We retain the pretrained FLUX.2
codec and transformer~\cite{BlackForestLabs2026Flux2}, and adapt the model with
LoRA~\cite{Hu2022LoRA} and the reference-token conditioning mechanism of
EasyControl~\cite{Zhang2025EasyControl}.

\paragraph{MRI intensity and radiomics.}
Histogram standardization, z-score normalization, fixed windows, and percentile
clipping can reduce MRI acquisition variation, but depend on anatomy, protocol,
foreground definition, and outliers~\cite{Nyul2000MRIStandardization,Reinhold2019MRIIntensityNormalization}. Preprocessing also affects radiomic
features~\cite{Zwanenburg2020IBSI}. Fr\'echet Radiomic Distance (FRD) compares
image cohorts in a standardized radiomic feature space~\cite{Konz2026FRD}. PEC
connects these representation concerns to conditional generation by enforcing a
shared coordinate at training time and estimating it from the source at
deployment.


\section{Method}
\label{sec:method}

\subsection{Latent VCE and Intensity Coordinates}

Let $x,y\in\mathbb{R}^{H\times W}$ denote a standardized pre-contrast breast
MR slice and its patient-specific peak post-contrast target. A latent VCE model
maps both images to the bounded input domain of an autoencoder and learns a
conditional generator $\hat y=G(x)$. The intensity mapping is therefore part of
the learned interface: it determines which MRI values the latent model can
represent and whether corresponding source and target values retain the same
meaning.

A fixed mapping uses one global interval for every patient. A low upper endpoint
clips enhancement, whereas a high endpoint assigns fewer 8-bit levels to the
densely occupied intensity range. Independent per-image normalization adapts to
each case but gives source and target pixels different coordinates. We instead
seek a mapping that is shared within each pair, adapts to the target enhancement
scale during training, and can be reconstructed from $x$ alone at inference.

\subsection{Predictive Enhancement Calibration}
\label{sec:pec}

Predictive Enhancement Calibration (PEC) defines the training endpoints as
\begin{equation}
    \ell(x)=\min_{p\in\Omega_f(x)}x_p,
    \qquad u(y)=Q_{99.99}(y),
    \label{eq:endpoints}
\end{equation}
where $\Omega_f(z)=\{p\mid z_p\text{ is finite}\}$ and $Q_{99.99}$ is likewise
computed over $\Omega_f(y)$ in the original field of view, including finite
background values. Endpoints are computed before canvas
padding; padded pixels are assigned $\ell(x)$ and therefore encode to zero. The
near-maximum percentile retains the enhancement tail while reducing sensitivity
to isolated outliers. Source and target are encoded with the same interval:
\begin{align}
 C_{\ell,u}(z)_p
 &=\left[\operatorname{clip}\left(
       \frac{z_p-\ell}{u-\ell},0,1\right)\right]^{1/\gamma},
       \quad \gamma=2.2,                                             \label{eq:encode}\\
 D_{\ell,u}(a)_p&=\ell+(u-\ell)a_p^{\gamma}.                         \label{eq:decode}
\end{align}
The gamma transform allocates more code levels to the lower and middle part of
the range before 8-bit quantization. The grayscale result is repeated over
three channels for the pretrained autoencoder. Sharing $[\ell,u]$ makes equal
encoded values comparable across the pair.

At inference, $u(y)$ is unavailable. We predict its log-span from 47 source
statistics---quantiles, tail means, top-$k$ summaries, threshold fractions,
moments, and image shape:
\begin{equation}
 r=\log\!\left(\max\{u(y)-\ell(x),\epsilon\}\right),\qquad
 \hat u(x)=\ell(x)+\exp f_{\phi}(s(x)).                              \label{eq:predict}
\end{equation}
The predictor $f_{\phi}$ is a four-block, 128-dimensional
FT-Transformer~\cite{Gorishniy2021FTTransformer}. It jointly estimates four
upper-tail percentiles with smooth-$L_1$ losses plus asymmetric and monotonicity
terms, observing target endpoints only as training labels. Deployment uses
$[\ell(x),\hat u(x)]$ for both source encoding and generated-image decoding.

\subsection{Reference-Conditioned Latent Flow}

We use the 9B FLUX.2 [klein] Base latent flow
transformer~\cite{BlackForestLabs2026Flux2}. The calibrated pre-contrast image
is encoded as reference tokens, while noisy target tokens follow the
flow-matching path. We adapt EasyControl's condition-token-only LoRA
principle~\cite{Zhang2025EasyControl} to the native FLUX.2 [klein]
reference-token key--value layout, using causal reference attention and
per-layer reference-key--value caching. The pretrained transformer,
autoencoder, and text-encoder weights are frozen; the rank-128 LoRA parameters
are trainable.

The flow-prediction loss emphasizes the lesion during training:
\begin{equation}
 \mathcal{L}=\frac{1}{|\Omega|}\sum_{p\in\Omega}
 \frac{1+\lambda m_p}{\operatorname{mean}(1+\lambda m)}
 \left\|v_{\theta,p}-v^{\star}_{p}\right\|_2^2,
 \qquad \lambda=4,                                                   \label{eq:roi_loss}
\end{equation}
where $m$ is the tumour mask downsampled to latent resolution. The mask affects
only the training loss; the inference path uses the pre-contrast image alone.

\section{Experimental Design}
\label{sec:experiments}

\subsection{Data and Evaluation}

MAMA-MIA contains 1,506 patients from multiple centres~\cite{Garrucho2025MAMAMIA}.
We use a fixed internal 100-patient development cohort, denoted \emph{MAMA100},
throughout model development and internal evaluation; the remaining 1,406
patients form the training pool. The target is constructed in two steps. First,
the post-contrast phase with the highest mean intensity over the three-dimensional
tumour ROI is selected for each patient. Second, the slice with the largest
tumour-mask area along the inferred through-plane axis is extracted from that
phase, together with the corresponding
pre-contrast slice~\cite{Osuala2026MAMASynth}.

The archived preprocessing applies one dataset-level z-score transform; PEC
follows it, so all endpoints are in z-score units. This global affine transform
leaves patient-specific dynamic-range variation and the bounded-codec mapping
problem. The predictor is fitted with the MAMA training patients and 80 of the
100 public Yunnan cases; 20 Yunnan cases are held out for predictor
selection~\cite{Zhang2023YunnanDCE}. Generator evaluation remains on MAMA100.

We report the eight benchmark metrics: MSE, LPIPS, tumour-region SSIM
(SSIM$_t$), FRD, contrast AUROC, tumour-ROI AUROC, Dice, and HD95. MSE is
computed directly in the common z-score domain. LPIPS clips both images to
$[-5,5]$ and maps them to $[-1,1]$ without per-image normalization before an
AlexNet backend; HD95 is reported in pixels. Lower is better for MSE, LPIPS,
FRD, and HD95. For metrics available per patient, we estimate paired 95\%
intervals with 10,000 bootstrap resamples; FRD and the AUROCs remain cohort-level
point estimates.

\subsection{Task-Aligned Lesion Sampling}
\label{sec:sampling}

Our earlier construction followed the multi-slice strategy of Ibarra et al.,
retaining tumour-positive slices plus a small fraction of nearby
negatives~\cite{Ibarra2025ConditionalDiffusion}. The present endpoint targets
peak enhancement at the largest lesion cross-section. We therefore rank axial
tumour-positive slices in each DUKE and ISPY2 patient by mask area and keep at
most eight (9,447 slices from 1,184 patients). Broader sampling increasingly
favoured peripheral, small-lesion slices and was less aligned with this endpoint.

\subsection{Comparisons and Implementation}

Target round trips compare two fixed intervals, target min--max, oracle PEC,
and source-predicted PEC, always using the same listed interval for encoding
and decoding. A percentile sweep separately measures tail clipping and FRD.

We compare three near-matched conditional models trained for 20 epochs on the
top-eight data with the 9B backbone, rank-128 LoRA, ROI loss, and $10^{-4}$
learning rate. Separate coordinates use the observed source maximum for encoding
and the predicted target endpoint for decoding. Fixed-wide uses $[-0.5,36]$ in
both directions; PEC uses $[\ell(x),\hat u(x)]$. Runs share the training seed,
comparable global batches, AdamW, and final-epoch selection. Inference uses a
$512\times512$ canvas, 30 steps, and the fixed prompt ``synthetic post-contrast
breast DCE-MRI slice, same anatomy, z-score normalized.''

\section{Results}
\label{sec:results}

\subsection{Intensity Calibration Before Generation}

In the floating-point sweep, moving the upper endpoint from $Q_{99.5}$ through
$Q_{99.9}$ and $Q_{99.95}$ to $Q_{99.99}$ reduces FRD from 10.598 through 3.526
and 1.923 to 0.090 as clipping falls from about 0.5\% to 0.01\%. Sparse bright
pixels therefore affect radiomic distance before generation.

Table~\ref{tab:roundtrip} confirms this representation effect. Oracle PEC has
the lowest direct FRD (1.280); target min--max has the lowest pointwise errors
and slightly lower post-VAE FRD, while the wide fixed interval remains strong.
Source-predicted PEC retains low MSE and high SSIM$_t$ but reaches FRD 9.186,
showing that global reconstruction metrics can miss the sparse radiomic tail.

\begin{table}[t]
\centering
\caption{Target round trips on MAMA100 ($n=100$). Encoding and decoding use the
same listed interval; FRD$_{\rm VAE}$ inserts the frozen autoencoder. Rows
marked oracle use target-derived bounds. Representation-only and synthesis
errors in Table~\ref{tab:coordinate} have different scales.}
\label{tab:roundtrip}
\scriptsize
\setlength{\tabcolsep}{3.0pt}
\begin{tabular}{lcccccc}
\toprule
Calibration & Target interval & MSE$\downarrow$ & LPIPS$\downarrow$ & SSIM$_t\uparrow$ & FRD$\downarrow$ & FRD$_{\rm VAE}\downarrow$ \\
\midrule
Fixed narrow & $[-0.5,18]$ & .01618 & .000093 & .9646 & 9.1249 & 9.9744 \\
Fixed wide & $[-0.5,36]$ & .00021 & .000177 & .9990 & 1.6627 & 3.1326 \\
Target min--max oracle & $[\min(y),\max(y)]$ & .00010 & .000044 & .9998 & 1.3958 & 2.7659 \\
PEC oracle & $[\ell(x),Q_{99.99}(y)]$ & .00026 & .000048 & .9996 & 1.2804 & 2.7925 \\
PEC source-predicted & $[\ell(x),\hat u(x)]$ & .00120 & .000328 & .9893 & 9.1862 & 9.6323 \\
\bottomrule
\end{tabular}
\end{table}

The predictor reaches endpoint MAE 1.371 and Pearson correlation 0.975, yet its
window clips a case-average 0.102\% of target pixels. A sensitivity control uses
oracle encoding followed by predicted decoding; it gives FRD 1.295 (2.797 with
the VAE) but larger pointwise errors. This separates endpoint scale error from
the predicted window's capacity to retain the tail.

The conditional experiment instead asks whether source encoding and output
decoding should share $\hat u(x)$; neither model accesses the target at inference.

\subsection{Shared Coordinates in Conditional Generation}

PEC gives the best point estimate for all eight metrics in
Table~\ref{tab:coordinate}. Against the matched fixed-wide generator, it reduces
MSE from 0.812 to 0.749 and FRD from 4.838 to 4.429; lesion-metric differences
are less certain.

\begin{table}[h]
\centering
\caption{Near-matched final-epoch synthesis on MAMA100 ($n=100$). Fixed-wide
uses $[-0.5,36]$; the other rows use a predicted endpoint. These z-score-domain
synthesis errors differ in scale from Table~\ref{tab:roundtrip}.}
\label{tab:coordinate}
\scriptsize
\setlength{\tabcolsep}{2.2pt}
\resizebox{\textwidth}{!}{%
\begin{tabular}{lcccccccc}
\toprule
Coordinate & MSE$\downarrow$ & LPIPS$\downarrow$ & SSIM$_t\uparrow$ & FRD$\downarrow$ & AUROC$_{con}\uparrow$ & AUROC$_{roi}\uparrow$ & Dice$\uparrow$ & HD95$\downarrow$ \\
\midrule
Separate source/target & .9096 & .1079 & .4837 & 4.8870 & .8962 & .4395 & .5845 & 92.07 \\
Shared fixed-wide & .8116 & .1042 & .4749 & 4.8381 & .9006 & .4237 & .5473 & 96.70 \\
Shared PEC & \textbf{.7493} & \textbf{.1017} & \textbf{.4859} & \textbf{4.4295} & \textbf{.9110} & \textbf{.4496} & \textbf{.5882} & \textbf{90.75} \\
\bottomrule
\end{tabular}}
\end{table}

Against fixed-wide, paired differences (PEC minus fixed-wide) are $-0.0623$ for
MSE (95\% CI: $-0.1108$ to $-0.0187$) and $-0.00249$ for LPIPS ($-0.00399$ to
$-0.00099$). The corresponding intervals for SSIM$_t$ ($-0.0047$ to $0.0267$),
Dice ($-0.0140$ to $0.0994$), and HD95 ($-37.30$ to $23.62$) include zero. FRD
and the two AUROCs are cohort-level point estimates.

\begin{figure}[t]
  \centering
  \includegraphics[width=0.85\textwidth]{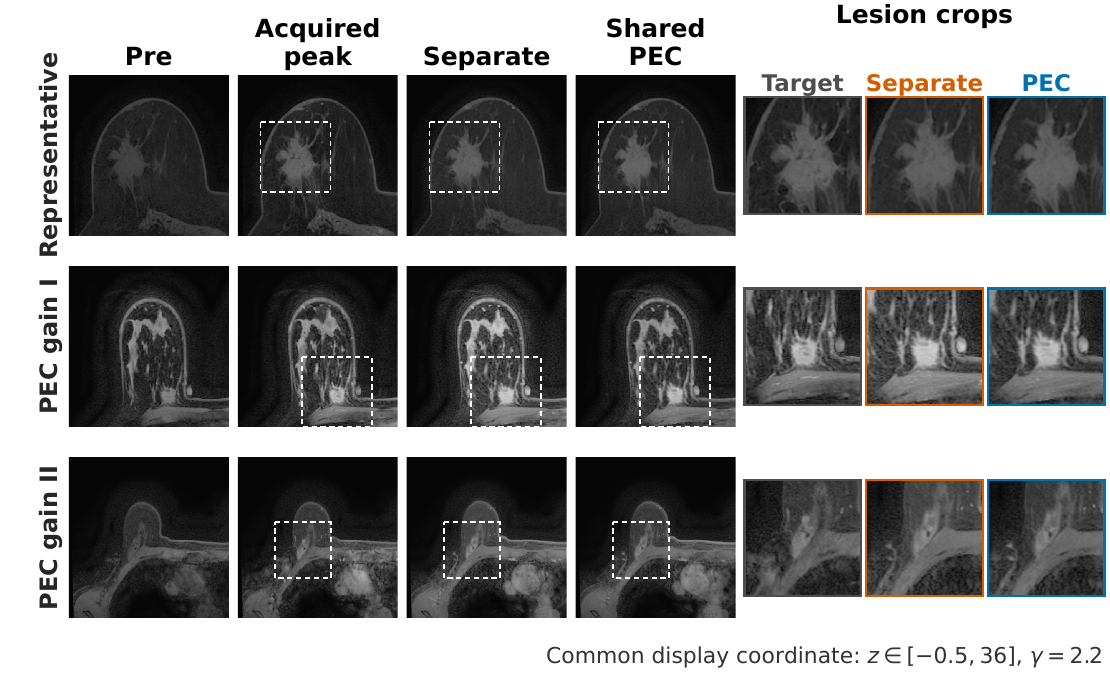}
  \caption{Qualitative MAMA100 comparison of a representative and two PEC-gain
  cases. Columns show the input, acquired target, separate-coordinate and PEC
  predictions, and matched crops; post-contrast panels share one coordinate.}
  \label{fig:qualitative}
\end{figure}

In Fig.~\ref{fig:qualitative}, the first case is the median of the middle
enhancement tertile. The other two have the best summed ranks among cases
improved in MSE, LPIPS, and SSIM$_t$; selection preceded rendering. Target and
predictions share display coordinate $[-0.5,36]$ and gamma 2.2.

\section{Discussion}
\label{sec:discussion}

Intensity mapping is not neutral: a misplaced upper endpoint changes FRD before
synthesis, although a wide fixed range is a strong control. Oracle and predicted
PEC separate coordinate capacity from source-only estimation accuracy. In
generation, sharing avoids an extra source--output scale change, while the
matched fixed-wide control supports case adaptation across all point estimates;
paired evidence is strongest for MSE and LPIPS.

High endpoint correlation can still hide tail clipping and code-allocation
errors in predicted-window round trips; the sensitivity control separates
window capacity from inverse-scale error.

Top-eight sampling targets peak-lesion cross-sections, unlike the broader
sampling of Ibarra et al.; dataset balance should follow the clinical endpoint.
A medical autoencoder is complementary but changes the latent basis.

FRD is cohort-level. MAMA100 informed development under archived
standardization. Train-only normalization, external cohorts, volumes, readers,
backbones, and LoRA, attention, loss, and sampling ablations remain future
tests.

\section{Conclusion}
\label{sec:conclusion}

PEC provides a shared, source-predicted intensity interface for latent breast
MRI VCE. It improves matched fixed-window synthesis while exposing tail
prediction as a remaining challenge.

\begin{credits}
\subsubsection{\ackname}
The authors gratefully acknowledge support from the Chongqing Science and
Technology Bureau: the 2024 Key Project of Technology Innovation and
Application Development, ``Research and Application of Precision Interactive
Integrated Medical Service Technology'' (Grant No.~CSTB2024TIAD-KPX0046); and
the Major Project of Technology Innovation and Application Development, ``Key
Technologies and Platform Development of Adaptive Multi-Task Large Medical
Models for Intelligent Diagnosis and Treatment'' (Grant
No.~CSTB2025TIAD-STX0029).

\subsubsection{\discintname}
The authors have no competing interests to declare that are relevant to the
content of this article.
\end{credits}

\clearpage
\bibliographystyle{splncs04}
\bibliography{references}

\end{document}